# Modified Binary Search Algorithm

Ankit R. Chadha
Dept. of Electronics & Telecommunication Engg.
Vidyalankar Institute of Technology
Mumbai, India

Rishikesh Misal
Department of Computer Engineering.
Vidyalankar Institute of Technology
Mumbai, India

Tanaya Mokashi
Department of Computer Engineering.
Vidyalankar Institute of Technology
Mumbai, India

## ABSTRACT
This paper proposes a modification to the traditional binary search algorithm in which it checks the presence of the input element with the middle element of the given set of elements at each iteration. Modified binary search algorithm optimizes the worst case of the binary search algorithm by comparing the input element with the first & last element of the data set along with the middle element and also checks the input number belongs to the range of numbers present in the given data set at each iteration there by reducing the time taken by the worst cases of binary search algorithm.

## 1. INTRODUCTION
Given the vast amount of data available, search algorithms are an inevitable part of programming. Searching broadly refers to the process of finding an item with specific properties in the given data set.

Tag Feedback based sorting algorithm for social search implements searching on the World Wide Web [1]. The Backwards Search Algorithm [2] is fundamental to retrieval of information in the full text model.

But the linear search and binary search algorithms form the base of many search applications. Linear search with its complexity of O(n) and Binary search with a complexity of O(log2 n) [3-4]have a high time complexity.

Hence, we have implemented the modified Binary Search algorithm with an aim of reducing execution time and increasing efficiency.

The rest of the paper is organized as follows. Section 2 explains the concept of Modified Binary Search. Section 3 and 4 describes the algorithm in detail. Section 5 shows the implementation of the algorithm for an example, followed by the performance analysis in Section 6. Finally, the conclusions are presented in Section 7.

## 2. Concept of Modified Binary Search
The concept of binary search uses only the middle element to check whether it matches with the input element. Due to this if the element is present at the 1st position it takes more time to search making it the worst case of the algorithm.

This paper suggests changes to the binary search algorithm which optimizes the worst cases of the traditional binary search. Modified also has the same pre-requisite that the given set of elements should be in sorted order.

Modified binary search not only checks whether the element is present in the middle index but it also checks whether it is present in the 1st and last position of the intermediate array at every iteration. Modified binary search is also optimized for the case when the input element does not exist which is out of range amongst the range present in the array that is, if the input number is less the 1st element or it is greater than the last element it certainly does not exist the given set of elements. Since it checks the lower index and higher index element in the same pass, it takes less number of passes to take a decision whether the element is present or not.

## 3. FLOW CHART
Below is the flow chart for the Modified binary search algorithm

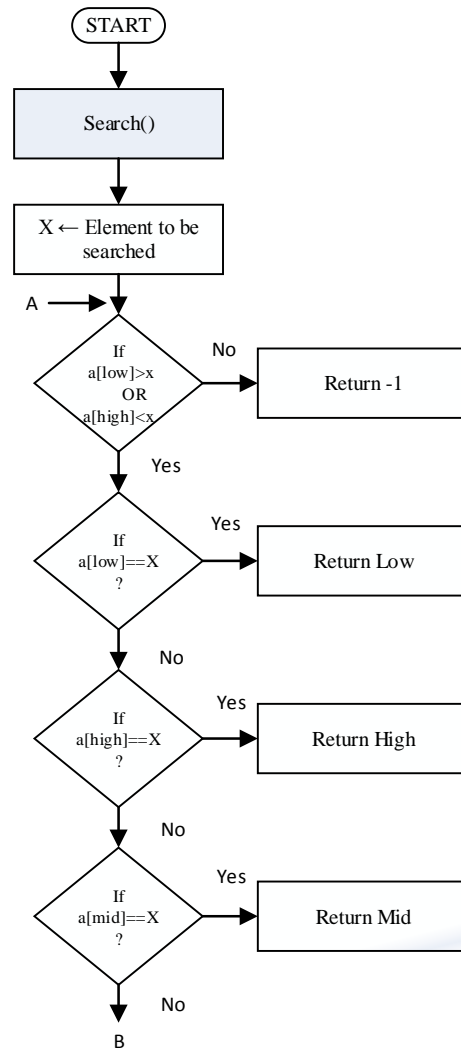

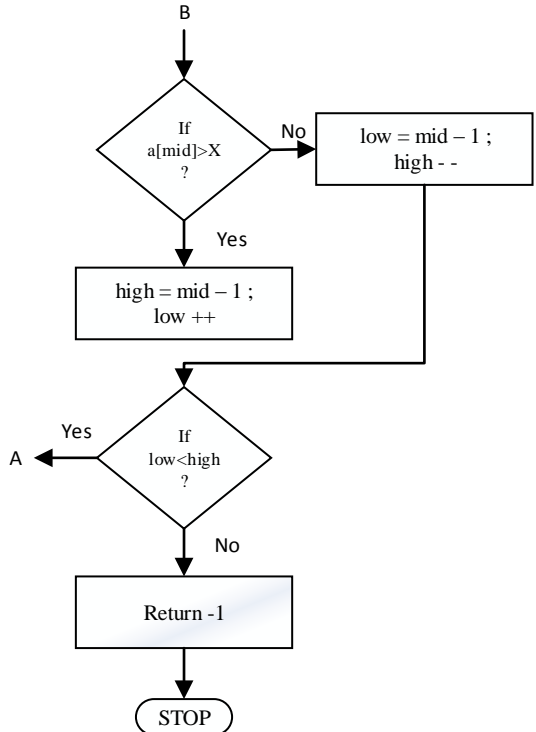

Terms used,
Search() – function call to check whether the given element is present or not

Low – the lowest index in the array
High- highest index in the array
Middle – middle element's index in the array
X – element to be searched

## 4. Pseudo Code
*X = number to be searched, a[] - elements array , 'n' total number of elements*

1. *low=0 , high= n-1*
2. *while(low<high)*
3. *mid=(low+high)/2*
4. *if(a[low]>X OR a[high]<X)*
5. *return -1*
6. *end if*
7. *if(a[low]==X)*
8. *return low*
9. *else if(a[high]==X)*
10. *return high*
11. *else*
12. *if(a[mid]==X)*
13. *return mid*
14. *else if(a[mid]>X)*
15. *high=mid-1*
16. *low++*
17. *else if(a[mid]<X)*
18. *low=mid+1*
19. *high - -*
20. *end if*
21. *end if*
22. *end while*
23. *return -1*

## 5. Example

| A[ ] | | | | | | | | | |
|---|---|---|---|---|---|---|---|---|---|
| 2 | 13 | 17 | 29 | 37 | 77 | 89 | 145 | 159 | 201 |
| 0 | 1 | 2 | 3 | 4 | 5 | 6 | 7 | 8 | 9 |

X - element to be searched

### 5.1 Element Present at 1$^{st}$ or Last position
X = 2
(i)Binary Search working:-
Pass 1 -
low=0 , high =9
mid=(low+high)/2
   = 4

A[mid] > X
high = mid-1

Pass 2 –
low =0 , high =3
mid=(low+high)/2
   = 1
A[mid] > X
high=mid-1

Pass 3 –
low=0, high=0
mid=(low+high)/2
   = 0
A[mid]==X

Element found at 1$^{st}$ position
Total Passes Required = 3

(ii)Modified Binary Search Algorithm

Pass 1 -
low=0 , high =9
mid=(low+high)/2
   = 4
A[low]==X
Element found at 1$^{st}$ Position
Total Passes Required = 1

### 5.2 Element present in second half of the array
X = 77
(i)Binary Search Algorithm
Pass 1 -
low=0 , high =9
mid=(low+high)/2
   = 4
A[mid] < X
low = mid +1

Pass 2 –
low =5 , high=9
mid=(low+high)/2
   =7
A[mid] > X
high = mid-1

Pass 3 –

low =5 , high=6
mid=(low+high)/2
     =5
A[mid]==X
Element Found at 6th position
Total Passes Required = 3

**(ii) Modified Binary Search Algorithm**

Pass 1 -
low=0 , high =9
mid=(low+high)/2
     = 4

Pass 4-
low=9, high=9
mid=(low+high)/2
     =9
A[mid]<X
low=mid+1

Since low>high
Element not found
Total Passes Required = 4

## 5.3 Element not present as is out of range among the input elements
X = 220
**(i)Binary Search Algorithm**
Pass 1 -
low=0 , high =9
mid=(low+high)/2
     = 4
A[mid] < X
low = mid +1

Pass 2 –
low =5 , high=9
mid=(low+high)/2
     =7
A[mid]<X
low=mid+1

Pass 3-
low=8, high=9
mid=(low+high)/2
     =8
A[mid]<X
low=mid+1

**(ii)Modified Binary Search Algorithm**

Pass 1 -
low=0 , high =9
mid=(low+high)/2
     = 4

A[high] < X
Element not found
Total Passes Required = 1

A[mid]< X
low=mid+1
high = high-1

Pass 2 –
low=5 , high =8
mid=(low+high)/2
     = 6
A[low] ==X
Element found at 6th position
Total Passes Required = 2

## 6. Performance Analysis
In the Below shown graphs x-axis represent the execution time taken by the two algorithms and the y-axis represents the number of input elements

### 6.1 Searching element in first half

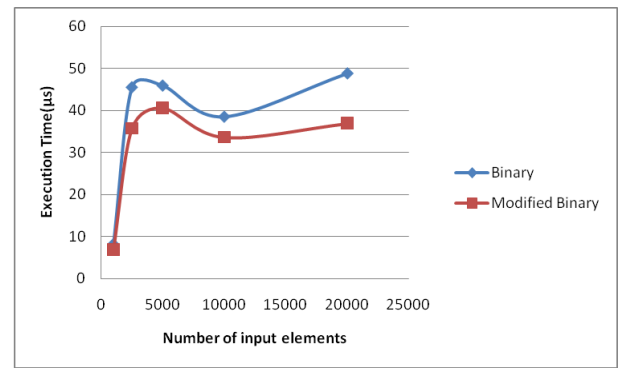

**Fig 1(a) : Searching Element in first half of data set**

The above graph shows time taken by each algorithm for the given number of elements to find an element which was present in the first half of the given set of elements.

### 6.2 Searching element at first/last position

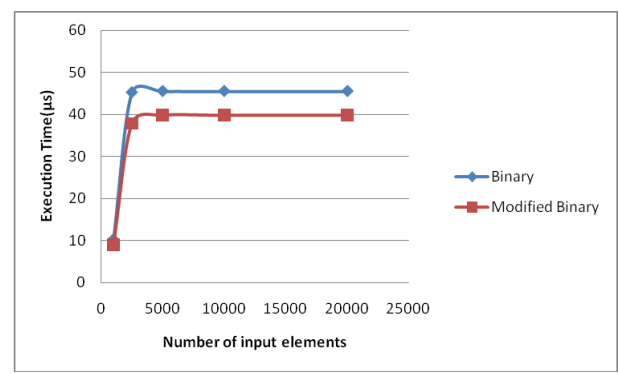

**Fig 1(b) : Searching First/Last Element in the data set**

The above graph shows time taken by each algorithm for the given number of elements to find an element which was present at the first/last position of the given set of elements.

## 6.3 Searching element does not exist and is out of range among the input elements

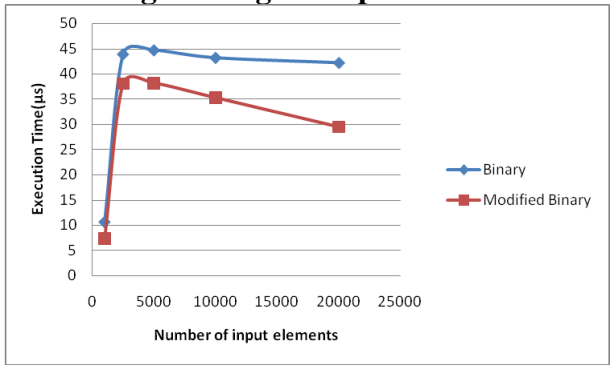

**Fig 1(c) : Searching Element that does not exist/ is out of range**

## 6.4 Searching element does not exist and is in of range among the input elements

The above graph shows time taken by each algorithm for the given number of elements to find an element which was not present in the given set of elements and was out of range.

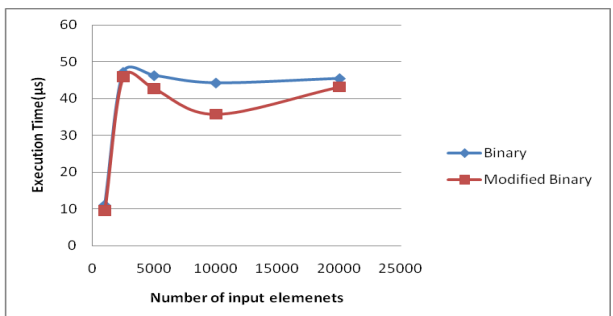

**Fig 1(d) : Searching Element that does not exist but is in range**

The above graph shows time taken by each algorithm for the given number of elements to find an element which was not present in the given set of elements and was in range.

The execution time shown in the above graphs was taken as the average of all the values considering different cases. Each time same element present at the same location was used to search for both the algorithms making it a fair comparison.

## 7. CONCLUSION

The modified binary search improves the execution time vastly over traditional binary search. The algorithm is comparatively more efficient as it eliminates unnecessary comparisons at the preliminary stage itself. This algorithm can even be extended to include the String domain.